\documentclass[conference]{IEEEtran}
\IEEEoverridecommandlockouts
\usepackage[
backend=biber,
style=numeric,
sorting=none
]{biblatex}
\usepackage{amsmath,amssymb,amsfonts}
\usepackage{algorithmic}
\usepackage{enumitem}
\usepackage{graphicx}
\usepackage{textcomp}
\usepackage[table]{xcolor}
\def\BibTeX{{\rm B\kern-.05em{\sc i\kern-.025em b}\kern-.08em
    T\kern-.1667em\lower.7ex\hbox{E}\kern-.125emX}}
\begin{document}

\title{Measuring the Concentration of Control in Contemporary Ethereum}

\author{\IEEEauthorblockN{Simon Brown}
\IEEEauthorblockA{\textit{ConsenSys Software Inc.}\\
simon.brown@consensys.net}
}

\maketitle

\begin{abstract}
Ethereum is undergoing significant changes to its architecture as it evolves.  These changes include its switch to PoS consensus and the introduction of significant infrastructural changes that do not require a change to the core protocol, but that fundamentally affect the way users interact with the network.  These changes represent an evolution toward a more modular architecture, in which there exists new exogenous vectors for centralization.  This paper builds on previous studies of decentralization of Ethereum to reflect these recent significant changes, and Ethereum's new modular paradigm.
\end{abstract}


\section{Introduction}

The contribution of this paper is to propose a model for measuring decentralization that accommodates structural changes in wider network topology over time.  As web3 and cryptocurrencies are a relatively nascent socio-technological innovation, they are in a phase of initial rapid innovation, in which the architecture and topology of the networks are evolving significantly.  This is a well understood phenomenon in technological innovation, which was documented as early as the 1960s \cite{rogers2010diffusion}, in which the innovation and adoption of new technologies develop in an “S-Curve” shape, involving compressed stages of very rapid innovation followed by a period where innovation plateaus for a time.  Ethereum is an example of a technology that is in the rapid innovation phase, in which there are significant changes to the topology of the overall network, both intrinsic and extrinsic to the core protocol.  Whereas previous research \cite{gochhayat2020measuring, lin2021measuring, karakostas2022sok} focused on measuring decentralization at the various layers of a vertical stack within a monolithic system, our model views Ethereum as a socio-technological ecosystem in which significant components of the network develop outside the core protocol.

The paper is organized as follows: in section II we deliver an overview of how Ethereum is evolving and the challenges faced when attempting to measure its level of decentralization.  In section III we outline the various dimensions that we propose to measure with our model, and our data sources.  In section IV we describe our methodology, including the various indices that are applied to our data.  In section V we deliver a breakdown of results, and we close with our conclusions in section VI.

\section{Background}

It can be argued that protocols that are built on top of the base layer of Ethereum do not pose a direct threat to the base layer itself, even when they are highly centralized, and should therefore not be a factor in quantifying the network's level of decentralization.  Once the base layer is sufficiently decentralized, any number of protocol designs can be implemented on top of it, and ideally the base layer should not be aware of them, or be adversely affected by them.  However, as Ethereum evolves, users increasingly interact with the network through abstracted layers of infrastructure that overlay the core protocol, and as such it can be conversely argued that such protocols could potentially affect the security and/or performance of the overall network under certain conditions, and should therefore be considered within a holistic model of measurement of the networks' level of decentralization. Our criteria for inclusion within our model is that the component being measured does not just serve a single use case or application, but is a protocol through which users interact with a substantial number of other dapps and protocols.

Any infrastructure that assumes a significant role in Ethereum can pose a centralization risk to the overall network based on two critical factors:
\begin{enumerate}[label=\alph*.]
\item the size of the infrastructure compared to the base layer, as measured in either the amount of base layer transactions that flow through the infrastructure and/or the Total Value Locked (TVL) compared to the base layer.
\item the potential effect on the base layer should the infrastructure be compromised or develop misaligned incentives, whether this effect is a level of effective degraded performance of the network, or an increased level of censorship.
\end{enumerate}

Ethereum is not a static ecosystem, and other innovations will likely assume a prominent role within the ecosystem in the future, e.g. EigenLayer \cite{eigenlayer2023}, DVT \cite{asgaonkar-2021}.  As such, any model that we develop should account for the changing topology of the ecosystem and allow us to incorporate new infrastructure into the model at a future date, while still being able to track the changes of effective decentralization over time.

\section{Selection of Data Points}

\subsection{Overview of Data Model}

We use as a base for our model the measurement of decentralization in blockchain first described by Balaji Srinivasan as the Minimum Nakamoto coefficient \cite{srinivasan2018}.  This model considers a blockchain network as being composed of a number of subsystems, which are important in terms of maintaining decentralization within an ecosystem, allowing it to remain resistant to capture by any one party or group.  Srinivasan describes any blockchain as being only as decentralized as the least decentralized subsystem, and his original model loosely defines a number of discrete subsystems to measure.

We have adapted Srinivasan's model to the contemporary PoS Ethereum topology and introduced several other dimensions that represent exogenous vectors for potential centralization.  Our model thus extends Srinivasan's original model from 6 dimensions to 12.  These dimensions of measurement are listed below, and are followed by a detailed explanation of the rationale for each.

\vspace{8pt}

\begin{itemize}
   \item \textbf{Based on original Nakamoto Coefficient subsystems:}
   \begin{itemize}
     \item Consensus nodes by client
     \item Consensus nodes by country
     \item Execution nodes by client
     \item Execution nodes by country
     \item Distribution of native asset by amount
     \item Amount staked by pool / staking service provider
   \end{itemize}
   \item \textbf{Metrics pertaining to PBS:}
   \begin{itemize}
       \item Blocks proposed by builder
       \item Blocks proposed by relay
   \end{itemize}
   \item \textbf{Metrics pertaining to Account Abstraction:}
   \begin{itemize}
       \item Number of user operations per bundler
       \item Number of wallets per deployer
   \end{itemize}
   \item \textbf{Miscellaneous Metrics:}
   \begin{itemize}
       \item Effective inflation rate adjusted for burn
       \item Percentage of total supply staked
       \item Layer 2 rollups by relative TVL
       \item Stablecoins by relative TVL
   \end{itemize}
 \end{itemize}

 \vspace{4pt}

\subsection{Metrics based on the Nakamoto Coefficient Subsystems}

We have adapted the Nakamoto Coefficient model through a number of modifications to the original model.  These changes include removing Mining Decentralization and Developer Decentralization. 

The ``Mining Decentralization'' metric is no longer relevant in PoS Ethereum and as such has been replaced by the ``Amount staked by pool'' metric, which measures the relevant share of the staked ETH by staking service provider. 

The ``Developer Decentralization'' metric is no longer an applicable metric for PoS Ethereum. The rationale for this change is the fact that nodes on the network run a number of different client implementations, each with its own distinct development team.  In this context, and considering a priori that developers are unique to each team, it is sufficient to measure the level of client diversity among nodes on the network rather than the relative contributions of individual developers. 

The ``Exchanges by Supply'' metric is not employed in our model as there is not a strong enough argument as to relevance of this metric to decentralization in Ethereum.

In terms of client diversity, it is also necessary to update the model to reflect the fact that Ethereum is now technically two merged blockchains that operate in unison, the Beacon Chain which handles consensus, and the execution layer, which is the P2P layer that gossips transactions and handles execution.  For this reason, the original ``Client Decentralization'', and ``Node Decentralization'' metrics have been replaced by ``Consensus / Execution nodes by client / country'' metrics. 

The only metric that has been retained in its original form is ``Distribution of native asset by amount'', which measures wealth inequality in terms of ownership of ETH.

\subsection{Metrics pertaining to Proposer Builder Separation}

Our model introduces two new metrics that pertain to Proposer Builder Separation (PBS), which are ``Blocks proposed by builder'' and ``Blocks proposed by relay'' respectively.

PBS is a network topology that has not been implemented at the protocol level, but has been implemented via the mev-boost middleware developed by Flashbots \cite{gosselin2021}, and which came online at the time of Ethereum's switch to PoS.

Fundamentally, PBS allows for the separation of concerns between block building and block proposing \cite{ethereum2023}, whereas currently the protocol assigns the responsibility of both to the validator.  Ethereum's PoS protocol requires validators to broadcast a valid block of transactions to the network when they are selected as a proposer.  As per the specification, validators will build a block locally by requesting their local execution client to collate pending transactions from the public P2P network. However, validators can install mev-boost and can request blocks from third party specialist block builders via public relays, instead of building one themselves \cite{ethereum2022}.

This has several benefits, from lowering the resource requirements for running a validator node, to reducing centralizing economics of MEV in staking pools \cite{buterin2021}.  However, it also introduces a number of other actors into Ethereum's infrastructure topology, i.e. block builders and relays, creating new vectors for potential centralization.

\begin{figure}[h]
\centering
\includegraphics[width=0.46\textwidth]{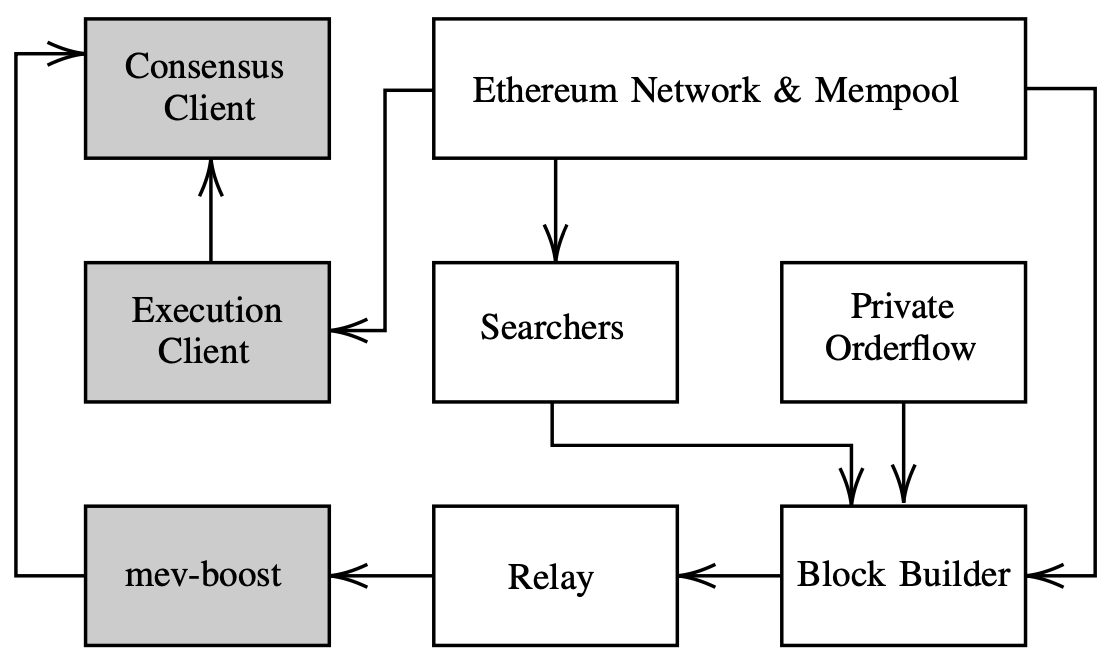}
\caption{High level mev-boost architecture}
\label{fig:mev-boost architecture}
\end{figure}

Our model applies a weighting to the PBS metrics when considering the measurement in the context of Ethereum's overall level of decentralization.  This is because any byzantine behavior of the mev-boost middleware will result in validator nodes falling back to local block production, thus preventing any safety or liveness fault within the core protocol \cite{hasu2022}.

However, it has been well documented \cite{labrys2022} that a number of prominent mev-boost builders and relays actively censor transactions according to specific criteria.  This effectively results in those transactions experiencing a potentially significant delay in being included in a block, (about 68\% longer than regular transactions according to Yang et al. \cite{yang2022sok}). 

This can effectively create a two-tier network with transactions associated with certain addresses becoming ``less privileged'' than other transactions.  Ironically the more transactions that are censored in this way, the harder it is to censor them, as block builders will need to bid higher than the combined value of those transactions in order to have their blocks proposed, resulting in an effective per-block fee for censoring transactions \cite{buterin2022}.  However, the higher the level of centralization within the block builder / relay infrastructure, the greater the risk for censorship, creating increased barriers to participation in the network for affected users.

\subsection{Metrics pertaining to Account Abstraction}

Our model introduces two metrics that pertain specifically to account abstraction, including ``Number of user operations per bundler'' and ``Number of wallets per deployer''. Account Abstraction has been a goal of Ethereum since its inception, and there have been a number of previous proposals that were not implemented \cite{john2023, wilson2020, dietrichs2020}, which all involved some change to the core protocol.  The breakthrough came with ``ERC-4337: Account Abstraction Using Alt Mempool'' \cite{buterin2021B}, which does not require a protocol change, but which introduces new roles within the ecosystem topology: bundlers and paymasters.

ERC-4337 specifies a specific transaction type called a user operation, or ``userop''.  User operations are submitted to bundlers, who batch them into a single transaction to a global entrypoint contract, which iterates over the userops in the batch, passing them to their respective smart contract wallets along with the userop's calldata for the contract wallet to execute (e.g. send ETH or call a function on some specific smart contract).

\begin{figure}[h]
\centering
\includegraphics[width=0.45\textwidth]{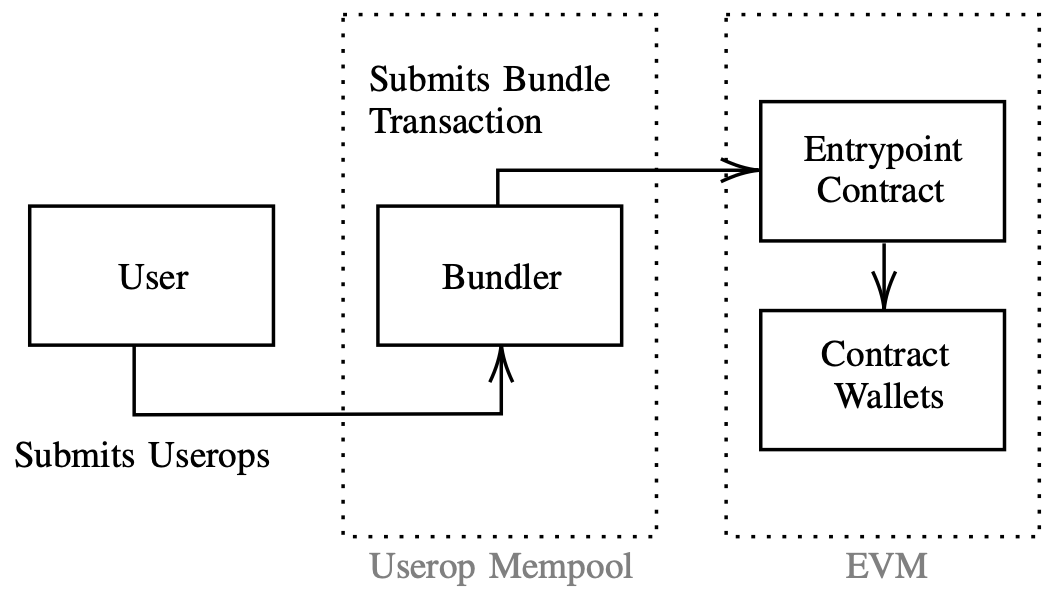}
\caption{ERC-4337 Account Abstraction Architecture}
\label{fig:4337 architecture}
\end{figure}

Our model weights these metrics lower than other metrics within the model, as the risk posed from centralization within this class of actors is lower than other parts of the infrastructure we have included.  Although bundlers can choose to censor specific transactions, the censored sender can simply decide to send their transaction directly to the entry-point contract, or to their smart contract wallet directly, if its design allows.  This represents a relatively weak form of censorship, but it does require some user sophistication in order to bypass. Over-centralization in this part of Ethereum's infrastructure can potentially lead to censorship risks, resulting in an effective two-tier network, with some addresses being less privileged than others in terms of their access to the network. We posit that this is a reasonable basis for including this metric in our model, albeit with an adjusted weighting.

\subsection{Layer 2 Rollups by Relative TVL}

Our model introduces a metric to measure L2 Rollups by TVL relative to the TVL of the base layer.  As Ethereum progresses through its ``rollup-centric roadmap'' \cite{buterin2020}, the TVL of L2 rollups as proportionate to the overall network becomes more significant within the composition of the ecosystem.

Our model applies a weighting to this metric, taking into account the extent of the risk that centralization within these protocols pose, i.e. they may not cause a safety or liveness fault in the underlying protocol, but may nevertheless potentially cause loss of funds or significant delays should they be compromised.

Consider as an example an L2 rollup with a centralized sequencer that experiences a significant liveness fault, in which users with funds on that network are no longer able to transact as they would under normal conditions.  In this scenario, it may be possible that the users can force a withdrawal through the L2's base layer smart contract bridge.  However, if this rollup contains a substantial number of user accounts, it may result in significant congestion on the base layer \cite{gorzny2022ideal}.  This would likely cause an increase in base fee and a prolonged delay in transaction inclusion.  Furthermore, in the case of tokens that are minted natively on an L2, it may not be possible to withdraw them to the base layer at all.

There is also a theoretical risk to Ethereum's underlying social consensus from having a single dominant rollup, which is described by Buterin \cite{buterin2023}, as forming a broad assumption within the ecosystem that "\textit{if there is a bug that causes funds to be stolen, the losses will be so large that the community will have no choice but to fork to recover the users' funds}".

\subsection{Miscellaneous Metrics}

As part of our model we measure the ``effective inflation rate adjusted for burn''.  This is an important metric for any PoS base layer protocol, including Ethereum.  A high rate of issuance of the network's native asset via validator rewards, has the effect of diluting the circulating supply, effectively decreasing the asset's value.  This incentivizes the network's users to stake the native asset in order to counteract the dilutionary effects of issuance, which forms a self-reinforcing cycle, leading to eventual hyperinflation even if that process takes a number of years.  Polynya describes a number of examples of this phenomenon that have been observed in practice \cite{polynya2022}. 

Our model thus incorporates a simple metric for measuring the inflation rate and adjusting it for the amount of ETH that is burned through the EIP-1559 mechanism, in which the base fee, which is adjusted for every block by the protocol itself, and that each transaction must pay at a minimum in order to be included in a block, is subsequently burned when a block is proposed.  This has the effect of  creating a negative issuance rate once transaction volume surpasses critical threshold, which will likely decrease the total supply over time \cite{liu2022empirical}.

We have also incorporated a closely related metric which is the ``percentage of total supply staked''.  This is directly related to the effective inflation rate metric with respect to maintaining an economic equilibrium between the issuance rate and circulating supply, allowing the asset to hold its value over time \cite{john2021equilibrium}.  It is worth pointing out that Ethereum's economics are designed to maintain this equilibrium by reducing issuance as more validators come online \cite{edgington2023}, which theoretically reduces the incentive to stake once the percentage of staked assets reaches a certain threshold. However, there is always the possibility that innovations such as EigenLayer may disrupt this equilibrium over time.

Another meaningful metric that we have introduced into our model is the measurement of ``Stablecoins by relative TVL on Ethereum''.  There exists both algorithmic stablecoins, (which are backed by a number of other assets and which rely on networks of decentralized oracles, and which are at least notionally decentralized by nature), and there also exists stablecoins which are backed largely by fiat deposits, and which are issued by a centralized authority.  The latter type of stablecoin is relevant to our model, insofar as while these stablecoins are not part of the infrastructure of the network, they are a significant part of the ecosystem with which users interact with other dapps.  They also pose a centralization risk in terms of censorship as there have been a number of  cases where stablecoins have been frozen from specific addresses \cite{wall2021}.

\section{Methodology}

The measurement of inequality in distributions is a well understood area of statistics that has found many applications in the fields of economics and social sciences. As our model aims to measure the level of decentralization across a number of different dimensions with different qualities, it is necessary to incorporate a number of different statistical measurements.

We use as a base measurement, the Gini index and the Herfindahl-Hirschman index. The Gini index is arguably the most widely used index with regards to measuring wealth inequality, and is well suited for measuring the distribution of a network's native asset, (i.e. ETH), while the Herfindahl-Hirschman index (HHI) is more often used for measuring the level of competition in specific industrial sectors, making it more suitable for measuring the degree of decentralization within the block builder market.  The base measurement is applied to each measurement dimension, but our results will reference one or the other based on the respective qualities of each.

The Shannon index is used to complement and cross reference any indicative result from our base measurement.  This is useful because the Shannon Index is based on a different approach to the Gini Index and the HHI.  The Atkinson Index is used to fine-tune the results from our base measurement by updating the index parameters to make it more sensitive to changes at different extremes of the distribution.  This can be useful if we want to qualify our results with more detail on any part of a distribution at certain times.

Tail ratios are used to inform any decisions about the parameterization of the Atkinson Index.  These include the P90:P10, and P50:P10 ratios, as well as the Palma ratio.  These ratios describe inequality within a distribution by comparing percentiles, for example, the P90 represents the level of resources allocated to $>$ 90\% of the population, while the P10 represents the level of resources allocated to $>$ 10\% of the population.  Our model adjusts the parameterization of the Atkinson index with respect the to the values of these tail ratios.

The Jensen-Shannon Divergence is used to track changes to the distribution over time, by comparing distributions of respective measurement dimensions between specific intervals, such as 30, 60 and 90 day intervals.

Our model also incorporates a master index in order to track the changes in each measurement dimension over discrete intervals.  This master index is an aggregate of other relevant indices, that is derived from calculating the normalized weighted geometric mean of the respective index's value across all other relevant metrics.

The indices that we have employed in our model are as listed below and are described in detail in the following sections.  Each index has characteristics and trade-offs associated with the underlying approach or model that they are based on.

\vspace{6pt}

\begin{center}
\begin{tabular}{|l|l|}
\hline
\textbf{Underlying Approach} & \textbf{Index} \\ \hline
Deviations model & Gini index \\ \hline
Combinatorics model & Herfindahl-Hirschman index \\ \hline
Entropy model & Shannon index \\ \hline
Social welfare model & Atkinson index \\ \hline
Tail ratios & Palma ratio, Pareto ratio \\ \hline
Divergence measures & Jensen-Shannon Divergence \\ \hline
Weighted geometric mean & Master Index \\ \hline
\end{tabular}
\end{center}

\vspace{2pt}

\subsection{Gini Index}

The Gini Index was developed by Corado Gini \cite{gini1936measure} as a mechanism for measuring inequality of income / wealth in a population, and is arguably the most commonly used measurement of inequality across a number of fields.   It is employed as the basis of the original Nakamoto Coefficient model, and has been used in several previous studies of decentralization in Bitcoin and/or Ethereum  \cite{sai2021characterizing, lee2021dq, gupta2018gini, kusmierz2022centralized, kwon2019impossibility, cong2023inclusion, gochhayat2020measuring, karakostas2022sok, zhang2022sok, campajola2022evolution}, which allows for some comparison with the results of previous studies.

The Gini Index is derived from the Lorenz curve \cite{lorenz1905methods}, which allows us to plot the individual shares of the distribution in relation to the overall total distribution. This becomes particularly useful for visualizing inequality within a distribution at a high level, and can also be very useful for comparing inequality between two distributions easily.

\begin{figure}[h]

\includegraphics[width=0.45\textwidth]{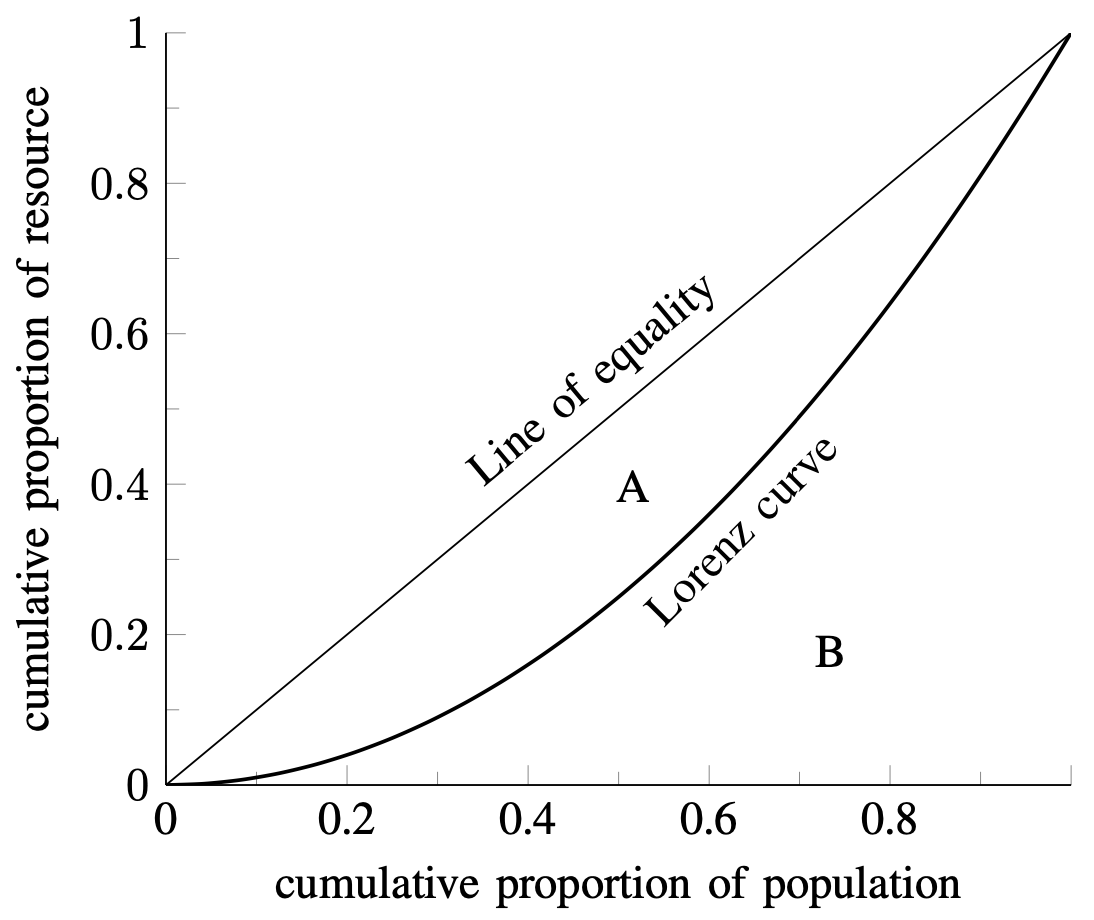}

\caption{Visualization of Lorenz Curve}
\label{fig:lorenz-curve-graph}
\end{figure}

There are a number of different methods of calculating the Gini Index (Tutberidze et al. describe four \cite{tutberidze2018measuring}), though most methods are based on the calculation of the Lorenz Curve of a population, whereby the Gini index is calculated as the area between the line of equality and the Lorenz curve, divided by the total area under the line of equality, i.e.: \(G=\frac{A}{A+B}\)

While the area under a curve is commonly calculated using the Newton-Leibnitz formula \cite{kalinski2016} , where $L(x)$ is the Lorenz curve.  It can also be approximated as the sum of the areas of a series of trapeziums, correlating in width to the unit interval being measured. 
\[G=1-2\int_{0}^{1}L\left( x \right)dx\]
\vspace{1pt}

While the calculation of the Gini Index using the Lorenz Curve is useful for visualizing the distribution and level of inequality on a chart, it is also possible to calculate directly using an approach that is based on the empirical mean difference of the values in the dataset \cite{gastwirth1972estimation}, which is the approach that we have implemented:

\[G=\frac{\sum_{i=1}^{n}\sum_{j=1}^{n}\left| x_i - x_j \right|}{2n^2\mu}\]

\vspace{4pt}

The Gini index gives us a value of between 0 and 1, where 0 indicates perfect equality of distribution of resources within the population, and 1 is total inequality, i.e. one single entity controls 100\% of the resources.

\subsection{Herfindahl-Hirschman Index}

The Herfindahl-Hirschman Index is commonly used as a measurement of competition within a certain industry sector. It has found applications in regulation, particularly with antitrust authorities \cite{usdoj2015}. It is calculated as the sum of the square of the percentage market share of each entity in a sector.

As the HHI is based on percentage shares of the market, it becomes close to zero for a market that has been commoditized, having a large number of participants with a relatively equal share.  Conversely, the HHI approaches 10,000 for a highly concentrated market, with 10,000, (or $1 \times 100^2 $), being a single entity monopoly.

Our model adapts the standard HHI by re-scaling it to make it comparable with other indices employed in the model, by dividing the HHI by $10^4$ so that it falls in the interval $0 <= \text{HHI} <= 1$. As such, the re-scaled HHI, denoted by $\theta$, is expressed via the following formula, where $n$ is the total number of participants in the market, $P$ is the total number of units produced in the entire market, and $p_i$ is the number of units consumed from $i^{th}$ participant.

\[\theta = \sum_{i=1}^{n} \frac{(p_{i}/P \cdot 100)^{2}}{10^{4}}
\]

As an example, in measuring concentration in the block building market, $P$ would be the total number of blocks produced in an interval,  $p_i$ would be the blocks proposed to the network that are built by the $i^{th}$ block builder.

The US DoJ generally classifies markets within three discrete categories \cite{usdoj2015}:

\vspace{8pt}

\begin{center}
\begin{tabular}{|l|l|}
\hline
Unconcentrated & $\text{HHI} < 1,500$  \\ \hline
Moderately Concentrated & $1,500 <= \text{HHI} <= 2,500$ \\ \hline
Highly Concentrated & $\text{HHI} > 2,500$ \\ \hline
\end{tabular}
\end{center}

\vspace{8pt}

Because the Herfindahl-Hirschman Index is commonly used for identifying and measuring the presence of monopolies in industry, it is more suited to the data points in our model that measure the infrastructure that is provided as a service or public good by a relatively small number of actors, as opposed to measuring the distribution of ownership or control of some asset.  This makes it particularly useful for measuring middleware such as block builders or relays, or ERC-4337 bundlers.

\subsection{Shannon Index}

The Shannon Index \cite{shannon1998mathematical} is part of a family of measurements that are derived from information theory and which are based on the concept of entropy.  Other measurements in this category include the Generalized Entropy measurement, and the Theil indices.  As the Shannon Index is based on a very different approach from either the Gini index or HHI, it is a useful measurement to cross-reference results against.

The Shannon index is a measure of the amount of entropy in a dataset.  It was intended to be used to measure the amount of information content in a signal, where information content can be considered as a measurement of the unexpectedness of a particular value occurring at a specific point in the signal, and entropy is the \textit{average} level of unexpectedness within a signal.  

Shannon gives examples of strings of characters, where the more characters and randomness there is, the lower the probability of predicting the next character in the string, and the more unexpected it is when that value occurs as predicted.  The less often the next character can be predicted correctly, the higher the entropy.

A basic example of this concept is a coin toss, where one party chooses heads, and where there is a 50\% chance of the expected value (i.e. heads) occurring as predicted.  As this has a relatively high probability, $p(heads) = 0.5$, coin tosses have low entropy. If we roll a die, the entropy increases as the probability of an expected value decreases, e.g. $p(6) = 0.16$.

Entropy as applied within the field of economics was systematized by Theil \cite{theil1967economics}, and has found applications in measuring inequality within a distribution of resources, and later within several studies of decentralization in cryptocurrencies \cite{zhang2022sok, gochhayat2020measuring, kusmierz2022centralized}.

The Shannon index is commonly expressed using the following formula, where $N_i$ is the frequency of a each value in the dataset divided by $N$, the total number of distinct values in the dataset, i.e. $N_i$ is the number of entities within the population that have a $i$ amount of resources, in proportion to $N$ total number entities in the population.

\[H' = - \sum_{i=1}^{N} \left(\frac{N_i}{N}\right) \log_{e} \left(\frac{N_i}{N}\right)\]

The Shannon index was designed for use with categorical data, as opposed to continuous data, for which the GE index is better suited \cite{tran2021harnessing}.   However, for the purposes of measuring distribution of native asset (i.e. ETH), we have applied the Shannon index to ranges of amounts of ETH.  In this context, $p_i$ is the proportion of the asset owned by the $i^{th}$ percentile of the population.

The Shannon index has a range between 0 and the logarithm of the number of categories in the dataset, i.e. $0<=H'<=log(n)$.  The more centralized a system becomes, the closer the Shannon index will be to zero \cite{kusmierz2022centralized}.

\subsection{Atkinson Index}

The Gini Index is useful as a base for calculating inequality, but it has limitations in describing the \textit{qualities} of inequality.  As the Gini Index is based on the ratio of total areas under the curve, it does not account for variance or skewness, and it also means that two different distributions can potentially have the same Gini index, which could potentially affect the tracking of changes over time.

For this reason, our model employs the Atkinson index \cite{atkinson1970measurement} as a further measure of decentralization, to allow us to cross-reference our Gini index against a measurement that can be fine-tuned to our requirements, and which can be used to capture any nuance in the distribution of different measurements.

The Atkinson index is based on the social-welfare approach, which infers the amount of resources that would need to be redistributed to achieve a certain level of equality.  The Atkinson index is calculated using the following formula:

\begin{align*}
A(\varepsilon) = 1 - \left(\frac{1}{N} \sum_{i=1}^{N} \left(\frac{y_i}{\mu}\right)^{1 - \varepsilon}\right)^{\frac{1}{1 - \varepsilon}}, \quad \quad \varepsilon \neq 1 \\
\quad \\
A\left( \varepsilon \right)=1-\frac{\prod_{i=1}^{N}\left( y_{i}^{\frac{1}{N}} \right)}{\mu}, \quad \quad \varepsilon=1
\end{align*}

\vspace{12pt}

where the parameters include:

\vspace{6pt}

\begin{center}
\begin{tabular}{|l|l|}
\hline
$\epsilon$ & inequality aversion parameter where $\epsilon>0$ \\ \hline
$n_{i}$ & number of people in the $i^{th}$ income group \\ \hline
N & total number of people \\ \hline
$y_{i}$ & average income of the $i^{th}$ income group \\ \hline
$\mu$ & average income of the total population \\ \hline
\end{tabular}
\end{center}

\vspace{6pt}

The Atkinson index is very closely linked to the generalized entropy index, and other related entropy based indices, such as the Theil index (i.e. GE(1)).  The Atkinson index results in a value between 0 and 1, and takes as a parameter $\epsilon$, which allows us to fine-tune the formula by increasing the value of $\epsilon$ in order to make the index more sensitive to changes at lower end of the distribution.  The value of $\epsilon$ is referred to as the inequality aversion, and the index yields a higher value when $\epsilon$ is given a value closer to 1.

\subsection{Tail Ratios}

According to Atkinson \cite{atkinson1970measurement}, the Gini index is affected by changes closer to the median of the distribution more than it is affected by changes at tail ends of the distribution.  In order to account for this characteristic of the Gini index, we employ a series of tail ratios to highlight any changes in the shape of the distribution that affect the lower and upper percentiles, and as a means to inform decisions about the inequality aversion parameter in our Atkinson Index.

Our model incorporates the Palma Ratio\cite{palma2011homogeneous}, which is the ratio of the share of resources allocated to the top 10\% of the distribution to the lower 40\% of the distribution.  Palma concludes that changes in the level of inequality tend to happen more at either end of the distribution, with the middle being affected less.

We complement the Palma ratio with other interdecile ratios that can be used to qualify the character of the inequality in the distribution, including the P90:P10, P50:P10 ratios.  Without loss of generality, Px represents the level of resources allocated to greater than x\% of the population, and is used to compare the gap between allocation of resources to two discrete percentiles within the distribution.

These inter-decile ratios are calculated using the linear interpolation method \cite{frost2023B} described below, where $P_x$ denotes the desired percentile, and $N$ is the population size, and $v$ represents the value at position $i$ in an ascending ordered dataset, where $i$ is calculated using the following formula:
\[ i = \frac{P_x\left( N+1 \right)}{100} \]
This gives us the rank position of the desired percentile, assuming the dataset is sorted in ascending order.  We then interpolate between the value at this rank position in the ordered dataset, and the value at the subsequent position, in order to attain each percentile in our desired inter-decile ratio:
\begin{align*}
P = \left( v_{\left\lfloor i+1 \right\rfloor} - v_{\left\lfloor i \right\rfloor}\right)(i \ \text{mod} \ 1) + v_{\left\lfloor i \right\rfloor} \ , & \ \ i \ \text{mod} \ 1 \neq 0 \\
P = v_i \ , & \ \ i \ \text{mod} \ 1 = 0
\end{align*}
By themselves these ratios are not useful measurements as the range within the results are not bounded, as with the Gini index or HHI. They can be useful to reference when deciding which end of a distribution warrants further scrutiny.

\subsection{Jensen-Shannon Divergence}

As we are interested in measuring the changes to levels of decentralization over time,  we employ a method described by Lin \cite{lin1991divergence} as the Jensen-Shannon Divergence, which is a method to measure of the level of similarity or divergence between two probability distributions.  The JSD is useful because it has a defined upper bound, which means it can be scaled for comparison with other index values.  In order to scale the JSD, we normalize it by dividing the result by the upper bound of the possible range of results, which in our model, is $log_2$.

We calculate the JSD between two intervals by comparing the distribution of each one across each measurement dimension respectively.  As such, we define $p$ and $q$  as two probability distributions, taken from two discrete intervals of a single metric, represented as vectors of sizes $n$ and $m$ respectively. The normalized JSD is then calculated as follows:

\[ D_{JS}(p||q) = \frac{1}{\log_2} \left( \frac{1}{2} D_{KL}(p||m) + \frac{1}{2} D_{KL}(q||m) \right) \]

\vspace{8pt}

where $m = \frac{1}{2} (p + q)$  is the average of $p$ and $q$, and $D_{KL}(p||m)$ and $D_{KL}(q||m)$ are the Kullback-Leibler divergences of $p$ and $q$  from $m$ respectively, defined as:

\vspace{6pt}

\[
D_{KL}(p||q) = \sum_{i=1}^{n} p(i) \log \frac{p(i)}{q(i)}
\]

\vspace{8pt}

It is important to note that both vectors are the same size, therefore if $n < m$,  we pad $p$ with zeros until it matches the size of $q$, or if $m < n$, we pad $q$ with zeros until it matches the size of $p$.  As our data is categorical in nature, we also align values each distribution such that $p$ and $q$ are ordered by category.  

The normalized JSD is calculated across each measurement dimension for both the 1 day, 30, 60 and 90 day intervals in order to track and quantify any changes in decentralization in specific subsystems over time.

\subsection{Deriving a Master Index}

As we are measuring Ethereum's level of decentralization over time, we need to account for its modular topology, in which significant components of its infrastructure will change.  For example, we might measure the concentration in the mev-boost relay market, but later decide to remove this metric as innovations like enshrined PBS make them redundant \cite{neuder2023}.

Our model calculates an aggregate set of indices across a number of dimensions of measurement, that can be used as further indicator of the relative level of decentralization in Ethereum as it changes over time.  Each relevant index, (i.e. Gini, HHI, Atkinson), contributes to an aggregate index that is derived from calculating the normalized weighted geometric mean of the respective index's value across all relevant metrics, i.e.:
\[ \gamma = \frac{{\left(\prod_{i=1}^{n} (\beta_i \times \omega_i)\cdot100\right)^{\frac{1}{n}} - \text{min}\left(\beta\right)}}{{(\text{max}\left(\beta\right) - \text{min}\left(\beta\right))\cdot10^{-2}}} \]

\vspace{8pt}

where \(\beta=\{\text{metric}_1, \text{metric}_2, ..., \text{metric}_n\}\) i.e. the set of relevant metrics, $\omega$ is the respective weighting for each metric, and $n=\left| \beta \right|$, the number of metrics being measured.

This master index can be used to see changes on a daily basis over a period of time and be used in conjunction with the normalized JSD, which measures changes at specific points of 30, 60, and 90 days.

The relevant metrics that are included in each aggregate index, along with their respective weightings, are listed below. The weightings are assigned based on the qualitative properties of the infrastructural component being measured. The actual weightings used in the formula are the percentage of each weighting relative to the sum of all weightings.

\vspace{6pt}
\begin{table}[h]
\normalsize
\begin{center}
\begin{tabular}{|l|c|}
\hline
\textbf{Metric} & \textbf{Weight} \\ \hline
Consensus nodes by client & 1 \\ \hline
Consensus nodes by country & 1 \\ \hline
Execution nodes by client & 1 \\ \hline
Execution nodes by country & 1 \\ \hline
Distribution of ETH by amount & 1 \\ \hline
Amount staked by Pool & 1 \\ \hline
Blocks proposed by builder & 0.7 \\ \hline
Blocks proposed by relay & 0.7 \\ \hline
Number of userops per bundler & 0.2 \\ \hline
Number of wallets per deployer & 0.2 \\ \hline
Layer 2 rollups by relative TVL & 0.5 \\ \hline
Stablecoins by relative TVL & 0.3 \\ \hline
\end{tabular}
\end{center}
\end{table}

This results in a series of aggregate indices for each interval within the time-range our sample data is taken from.  The aggregate indices relate to the Gini, Atkinson, Normalized HHI and Normalized Shannon indices.  This should allow us to track the changes in decentralization over time, while allowing us to examine an individual index at specific intervals in order to further understand any changes that occur.

\section{Results}

In this section we discuss the results of the application of our model to a sample that was recorded at 24 hour intervals across a period of 90 days, between the 23\textsuperscript{rd} May 2023 to the 23\textsuperscript{rd} of August 2023.  We discuss how the results demonstrate our hypothesis that decentralization is a dynamic quality that changes over time, rather than a static quality that is maintained in continuous equilibrium.

\subsection{Discussion of Results}

As can be seen from the table below, different indices give divergent results when applied to the same data.  This highlights the need to leverage more than one approach to measuring decentralization, and also highlights the fact that different indices are more suitable for applying to certain dimensions of measurements over others.

\begin{figure}[ht]
\begin{center}
\includegraphics[width=0.49\textwidth]{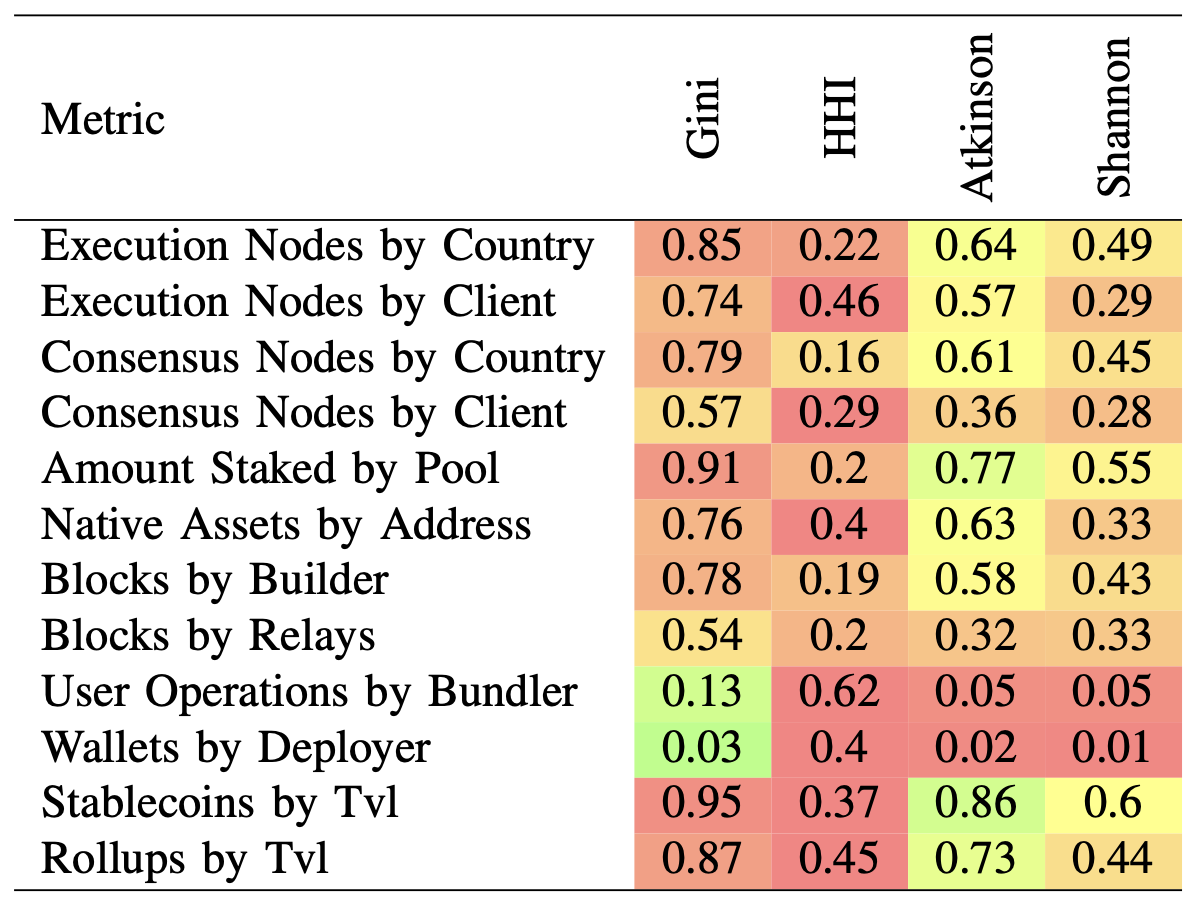}
\end{center}
\caption{90 day averages across all metrics}
\label{fig:average indices}
\end{figure}

The values in the table in figure \ref{fig:average indices} are color coded for legibility, with colors closer to red indicating higher levels of concentration, and colors closer to green indicating higher levels of decentralization. We observe that the index values for several metrics diverge significantly, particularly between the Gini index and HHI.  In most cases this is expected, since the HHI is more focused on measuring concentration at the upper end of a distribution. Consider, without loss of generality, a distribution with a large number entities that have a low share of resources. In this distribution the Gini index will increase significantly, while the HHI will remain largely unaffected.  If the same distribution contains a small number of entities that control nearly all the resources but that each have an equal share, the HHI will be relatively low, but will increase as the share between these controlling entities becomes less equal. 

We observe this phenomenon clearly in several metrics in the results table, for example in the Amount Staked by Pool metric, where 2 entities control ~60\% of the market\cite{hildobby2023} (Lido controls ~31\% while solo stakers control ~29\%), where the distribution of these two entities is almost equal, and the distribution between entities controlling the other ~40\% of the market is also relatively equal, which results in a re-scaled HHI of 0.2, indicating only moderate concentration, while the Gini index gives a value of 0.91, which more accurately reflects the level of concentration in this area.

While the Gini index and the HHI are each better suited to specific dimensions of measurement, there can often be nuances that they fail to capture by themselves.  An illustrative given by Buterin \cite{buterin2022b} is where there are two hypothetical societies with profound levels of inequality, one in which half the population equally shares all the resources while the other half has none, and the other in which one person has half of all the resources, everyone else equally shares the remaining half.  In Buterin's example, both distributions would result in the same Gini index value, despite having radically different characteristics.

These limitations form the rationale behind in the inclusion of the Shannon index within our model.  The Shannon index is sensitive to both size of the distribution and the diversity of different values in the distribution, and is therefore able to highlight differences between distributions that the Gini index does not capture.

This is meaningful in considering the level of concentration in areas where the distribution is smaller, for example with "Stablecoins by Tvl".  In the result for this metrics, the Gini index is very high, indicating high concentration, but the Shannon Index shows a relatively low concentration, and this is a reflection of the fact there are fewer stablecoins, and hence less chance for a decentralized market to emerge.

As we can see there is broad agreement between the Atkinson and Shannon indices, and this is not surprising considering they are both based on the entropy model.  However, the Atkinson index is not used by our model except for certain specific dimensions of measurements, because of its non-standard inequality aversion parameter, which makes it unsuitable for comparison with other studies. It can be useful however, in fine tuning certain measurements when needed.

The observations of the recorded results have been organized into the following sections which broadly correspond to the categories of metrics as described in section I.

\subsection{Results based on original Nakamoto Coefficient subsystem selection}

\subsubsection{Network Nodes}
\begin{figure}[ht]
\begin{center}
\includegraphics[width=0.46\textwidth]{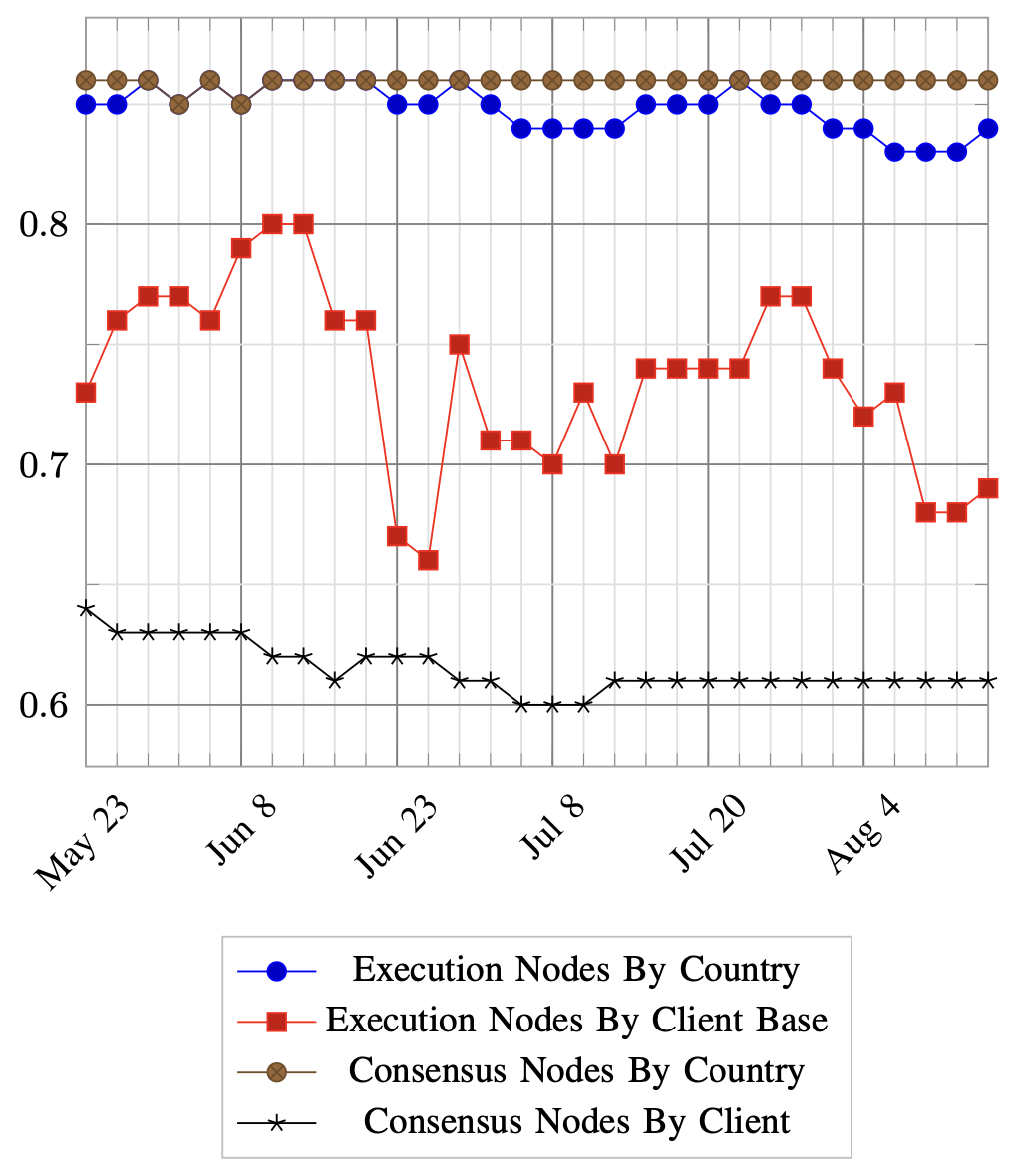}
\end{center}
\caption{Gini Indices for Network Nodes}
\label{fig:gini index network nodes}
\end{figure}

When examining the data recorded across the dimensions of measurement that are based on original Nakamoto Coefficient subsystem selection, we first applied the Gini index to the metrics pertaining to network nodes.  The results of this analysis is displayed in figure \ref{fig:gini index network nodes}, which plots the Gini index of these metrics across our sample time period.

We observe a relatively high Gini index value for both Execution Nodes by Country and Consensus Nodes by Country, which both maintain an average Gini index of 0.85 to 0.79.  When examining the data more closely, this was found to be caused by an out-sized number of nodes that are situated in the USA, having  ~34\% of Consensus nodes and ~44\% of Execution nodes.  Both numbers are significant in terms of having crossed a key consensus threshold of 33\%, which is enough to cause disruption to liveness under adverse conditions.

We observe that measuring Consensus Nodes by Client yields an average Gini index of 0.57, which is a reasonably safe level of client diversity, while Execution Nodes by Client shows a surprisingly dynamic Gini index, which changes between 0.66 to 0.81.  Reaching 0.8 is unhealthy for the network, as a bug in a single client version can have consequences for the network, as occurred in May 2023 when a bug in a consensus client caused a lack of finality for a period of time \cite{offchainlabs2023}.

We also refer to the HHI as measurement of client diversity, which may be a more more accurate measurement than the Gini index, due to the fact that  the population size of the distribution.  In both cases the HHI suggest a highly concentrated distribution, and therefore a lack of healthy diversity, with a slight over-concentration in consensus nodes (mainly driven by prism and Lighthouse \cite{abu2023}), and a pronounce over-concentration in execution nodes (driven by Geth).

\subsubsection{Amount Staked by Staking Pool}
In examining the \textbf{Amount Staked by Staking Pool}, we observe that the level of concentration stays relatively static with a Gini index of \textbf{90\%} to \textbf{91\%} during the sample period.  This suggests a very high level of centralization amount staking pools, however upon scrutinizing the the underlying data, we observe that ~30\% of the total ETH staked on the network is controlled by a single entity, which is Lido.  While this seems concerning at first, it is to be noted that Lido itself is a DAO, and as such is a decentralized entity.  Lido commissions 29 separate independent node operators to manage their validators, with a relatively even distribution of stake between each node operator, as documented on the VaNoM portal\cite{lido2023}.

Rather than accounting for Lido's decentralized quality by breaking down the composition of its node operators within the original dataset, we leverage the Atkinson index, by modifying the inequality aversion parameter to account for its level of decentralization.  To do this we multiply the standard inequality parameter of $0.5$ by a weighting derived from Lido's market share, denoted by $\omega$, i.e.:

\[\alpha^\prime = \alpha \cdot \left( 1 - \omega \right) = 0.5 \cdot \left( 1 - 0.3 \right) = 0.35\]

By using an inequality aversion parameter of $\alpha^\prime=0.35$ we obtain an Atkinson index of $0.6$ to $0.61$ which remains relatively static through the sample period, and which represents a much lower level of centralization than suggested by applying the Gini index.  The value of the Atkinson index indicates a moderate level of centralization, which is likely caused by a number of entities with larger of market shares, such as Coinbase, which has a share of ~10\%.

\subsubsection{Distribution of Native asset by Amount}

We observe that there is an indicative level of centralization with regards to the distribution of the native asset, ETH.  The most widely used index for measuring the distribution of wealth within an economic system is the Gini index, which gives an average value of 0.76 for the distribution of ETH.

While this represents a relatively high level of centralization with regards to control of ETH, we must make some consideration to the qualities of the entities at the top end of the distribution, that control large amounts of the asset, which could include the smart contracts of DeFi dapps, bridges etc..  One previous study by Glassnode \cite{checkmate2021} reported that 22.8\% of all ETH is stored in smart contracts.  This suggests that actual level of centralization in terms of ETH ownership is considerably less than is suggested through the  application of the Gini index.  While a portion of the smart contracts that hold ETH are wallets of private individuals, rather than dapps, the relative proportions of each is a question for further study.

\subsection{Metrics pertaining to PBS}

\begin{figure}[ht]
\includegraphics[width=0.45\textwidth]{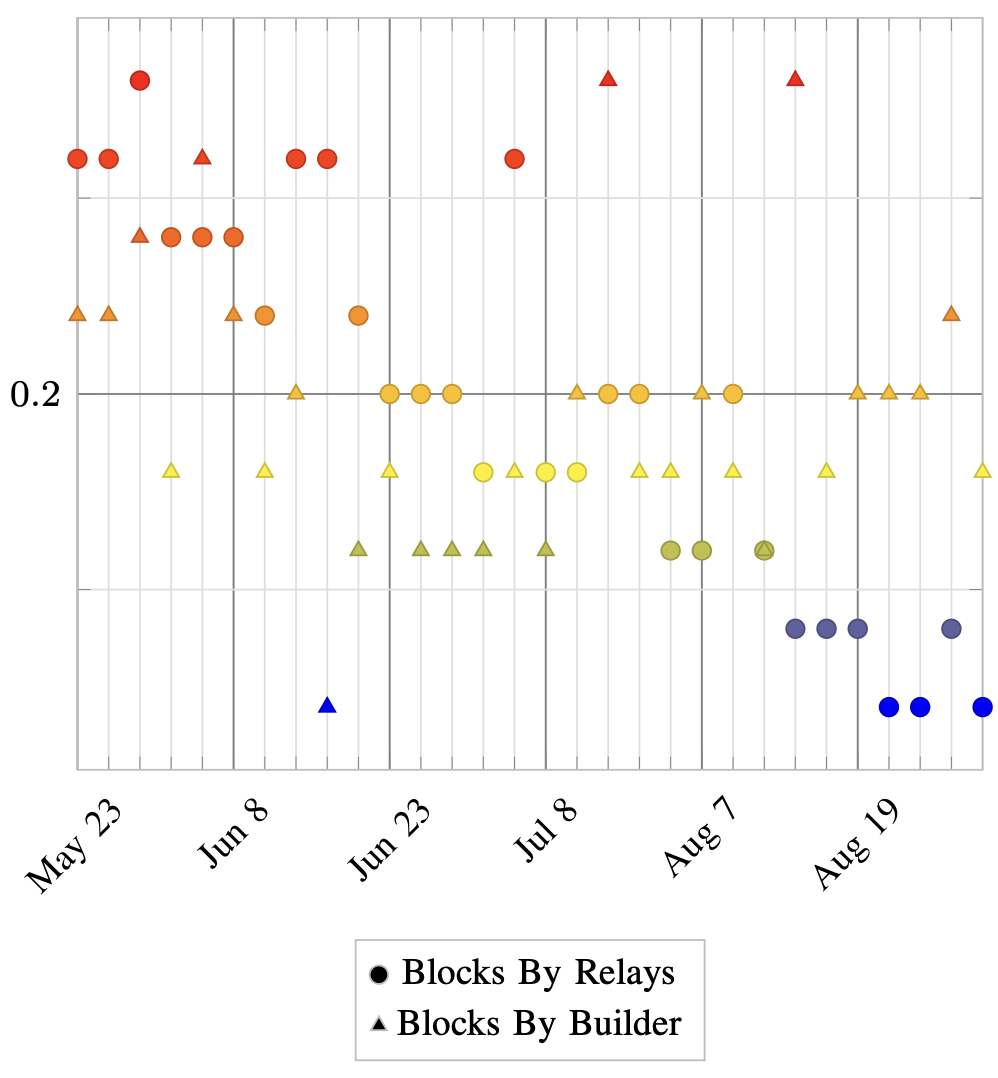}
\caption{HHI of Block Builders and Relays}
\label{fig:HHI block builder relays}
\end{figure}

In analyzing the data with regards to the metrics pertaining to Proposer-Builder-Separation, we examined both the metrics for Blocks proposed by Builder and the Blocks proposed by Relay.  For these metrics we applied the Herfindahl-Hirschman index as a more suitable index, due to the fact that the builder and relay space is naturally more concentrated, and comprised of a number of known entities, making it similar in some ways to more traditional economic sectors (as opposed to ~500,000 pseudonymous validators for instance).  The results of our analysis is visualized in figure \ref{fig:HHI block builder relays}.

As can be observed from the visualization, the market concentration ranges from a re-scaled HHI of ~0.16 (or 1,600), to ~0.24 (or 2,400) for relays, whereas the builders display a HHI that ranges from ~0.15 (1,500) to ~0.26 (2,600).

To refer to the description of the HHI in section II, we recall that the US DoJ regards any sector with a HHI of less than 1,500 as being unconcentrated, whereas anything above 2,500 as being highly concentrated. 

Using these broad categories as a reference, we observe that the block builder and relay markets range from almost unconcentrated to highly concentrated within the sample period.  We also observe that the level of concentration changes with significant relevance for the above categorization, even within a 90 day period.

When examining the Jensen-Shannon Divergence between the first and last 24 hour intervals in the dataset for both relevant metrics, we see a relatively high value compared to all other metrics, which indicates that there are significant changes in the number of blocks proposed by individual builders and relays, as well as overall changes to level of concentration within the market.  This is a useful index to cross-reference as it is capable of capturing nuance that other indices sometimes fail to capture, for example in cases where two or more entities can frequently reverse their respective market shares, with each claiming an outsized portion of the market for a period of time.  In this scenario, the overall level of inequality and centralization remains static but the relative share of individual participants can change radically.  This is a quality that the JSD index can help to capture.

\vspace{8pt}

\begin{center}
\begin{tabular*}{\linewidth}{@{\extracolsep{\fill}} llllll }
\hline
\textbf{Metric} & \textbf{JSD} \\ \hline
Amount Staked by Pool & 0.0150451 \\
Execution Nodes by Country & 0.0073133 \\
Execution Nodes by ClientBase & 0.0118433 \\
Consensus Nodes by Country & 0.0058399 \\
Consensus Nodes by Client & 0.0077753 \\
\textbf{Blocks by Relays} & \textbf{0.1324419} \\
\textbf{Blocks by Builder} & \textbf{0.2201886} \\
Stablecoins by Tvl & 0.0103413 \\
Rollups by Tvl & 0.1058467 \\
Native Assets by Address & 0.0005811 \\ \hline
\end{tabular*}
\end{center}

\vspace{8pt}

It is worth noting that over a longer time-frame, our model should be able to account for changes to the ecosystem from newer innovations and architectural changes.  These include enshrined PBS or the introduction of distributed block building, and there are a number of such proposals under development currently.

\subsection{Miscellaneous Metrics}

\subsubsection{Rollups by TVL}

We observe a pronounce level of centralization when we examine Rollups by TVL, with a 90 day average Gini index value of 0.87, and a re-scaled HHI of 0.45.  Both of these values make sense when we observe the underlying data, in which a single rollup has 54.3\% of TVL across all rollups, which is Arbitrum One (down from 64.5\% at the start of the sample period).  This is closely followed by Optimism that has a TVL of 25.9\%.  There are 12 rollups with a TVL share of less than 1\%, and 7 with a TVL of more than 1\%, the latter group exhibiting significant variance, from which we would expect a high HHI value.

Interestingly, the one other metric that also show a high JSD value for the sample period is Rollups By TVL.  This could be explained by the launch of the Base network on August 9\textsuperscript{th} which significantly altered the relevant market share of prominent rollups.

We observe that the level of centralization in the rollup market at the time the data was collected, is at an unhealthy level for the Ethereum ecosystem.  The TVL across all rollups is ~8.1B USD, which is approximately 0.3\% of the 26.562B USD TVL of Ethereum \cite{defillama}, and of which one single rollup accounts for 0.2\% of Ethereum's TVL.

It is worth noting again that these figures will more than likely change over time as a number of zk-EVM rollups will enter the market, which will dilute the marketplace and will likely affect market concentration significantly, underpinning the need for ongoing measurement of decentralization over time.

\subsubsection{Stablecoins by TVL}

We observe the highest average value for the Gini index in the ``Stablecoins by TVL'' metric, which also has a markedly high HHI value.  This suggests a very high level of concentration within the stablecoin market, which is obvious when we examine the underlying data, that shows that just two stablecoins comprise over 80\% of the market, i.e. USDC and USDT. As both stablecoins are issued by corporations, this represents a significant area of centralization within Ethereum.  Stablecoins are used by many people to interact with a wide array of dapps, and both USDC and USDT are vulnerable to censorship, and have previously frozen token holders funds on request from authorities \cite{haqshanas2022} \cite{de2022}.

According to DefiLlama \cite{defillama}, the combined market cap for stablecoins on Ethereum is ~9.2B USD, which represents a third of the TVL on Ethereum.  This could arguably be the weakest point of Ethereum's overall decentralization.

\subsubsection{Effective inflation rate adjusted for burn}

As discussed in section III, the Effective inflation rate adjusted for burn is an important metric to measure, as a high inflation rate can undermine the security of the network by diluting supply, and the devaluing the asset, potentially leading to misaligned incentives for the validators.

Since the introduction of EIP-1559, which introduces the concept of the "base fee", which is set deterministically via protocol consensus and which is burned in every block, Ethereum's effective issuance rate becomes negative when the network usage / gas price crosses a certain threshold \cite{ethereum2023-2}.

According to data from the Ultra Sound Money resource \cite{ultrasoundmoney2023}, the effective level of inflation of ETH since the merge has been -0.94\%, and ETH is projected to continue to be a deflationary asset in the long term, which is positive for the network's overall security and level of decentralization.

\subsubsection{Percentage of total supply staked}

The percentage of total supply staked is related to the effective inflation rate for the same reason as previously outlined, in that a high issuance has a dilutionary effect on the circulating supply, which incentives asset holders to stake in order to counteract the dilution, resulting in a feedback cycle that can undermine the network's security.  Ethereum’s economics are designed to reduce issuance as more validators come online [26], which theoretically reduces the incentive to stake once the percentage of staked assets reaches a certain threshold.  Currently the amount of ETH staked is 24,932,109 ETH according to beaconcha.in \cite{beaconchain2023}, out of 120,218,472 ETH total supply according to Etherscan \cite{etherscan2023}, resulting in a ratio of ~20\%, which is well within the bounds of what is  a safe level of total supply staked.

\subsection{Metrics pertaining to Account Abstraction}

The metrics that pertain to Account Abstraction, which are User Operations by Bundler and Wallets by Deployer are both striking insofar as they have especially low averages for the Gini index but very high values for the HHI values.  This indicates that there is an issue with the dataset that warrants further scrutiny.  Upon investigation we observe that the 90 day Gini indices for User Operations by Bundler has a range of 0 to 1, with a median of 0.6 and a standard deviation of 0.36.  This highly unusual range and variance can be attributed to the nascent nature of the account abstraction space, and indicates that the infrastructure has yet to see significant maturity or adoption.

Upon inspecting the underlying data, we observe that the maximum number of user operations processed in a single day during the sample period is only 301.  At this nascent stage, there is significant market concentration, with one specific bundler having 76\% market share out of 13 total active bundlers.

It is worth noting however that this data looks at Ethereum mainnet itself, and does not consider any of the L2 rollups, on which there may be a higher level of adoption of account abstraction infrastructure, and which is an area for future study.

The Wallets by Deployer metric shows very similar characteristics to the User Operations by Bundler and follows much the same pattern of distribution throughout the sample period, with one specific deployer having an 87\% market share.

The various metrics outlined in this paper will be useful in tracking the changes in decentralization over time as the account abstraction space matures and continues to see more adoption.  The influence of erc-4337 bundlers on the effective level of decentralization of the network as a whole is somewhat limited, so even a moderately high level of centralization is acceptable.

\subsection{Master Indices}

For every index that we track, we derive a master index that is a aggregate value derived from the value of each dimension of measurement for that index.  This means that we can derive a single value for every day in the sample period, rather than a 12 separate values for each day, (i.e. one for each dimension of measurement).

\begin{figure}[ht]
    \centering
    \includegraphics[width=0.49\textwidth]{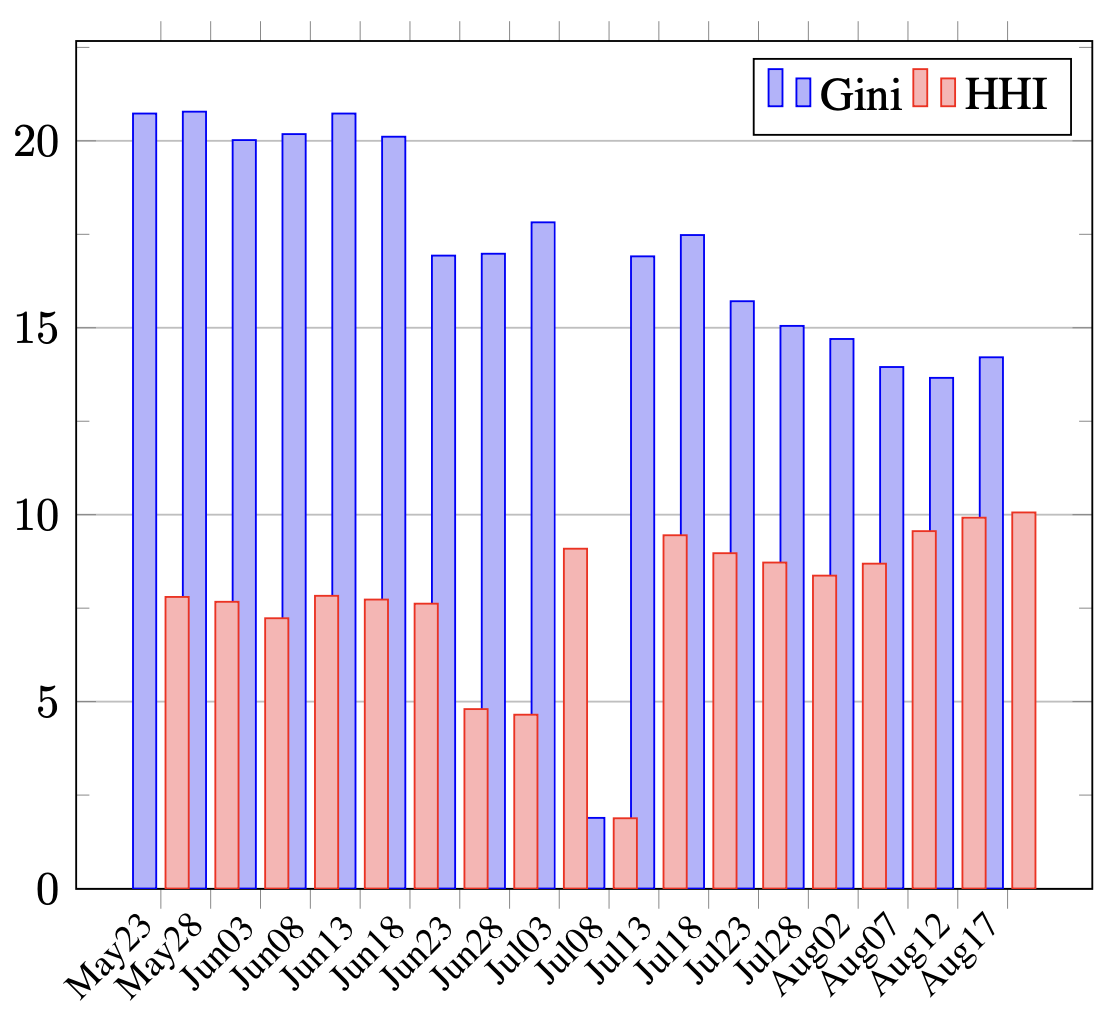}
    \caption{Master Indices of Gini Index and Herfindahl-Hirschman Index}
    \label{fig:masterindices}
\end{figure}

We are specifically interested in the master index for both the Gini Index and Herfindahl-Hirschman Index for each day in the sample period.  We have sub-sampled 18 data points from the 90 day period and have charted them in figure \ref{fig:masterindices}.

The master index establishes a measurement that can be referenced to itself as it changes over time, in order to derive a crude high-level indicator of the overall effective level of measurement of the network, and in theory we should be maintain a single relative measurement even as the dimensions of measure change over time.

As can be seen from the sample above, we can see that the overall effective level of decentralization changes over time, and this is a reflection of the changes within the various dimensions of measurement, specifically the Gini index and HHI index in this case.  Overall we can see a trend in a reduction in the value of the Gini across all dimensions of measurement from ~21\% to ~14\%, conversely the trend in the HHI for across dimensions of measurement increases slightly from 7.5\% to 10\%. Note that the anomalous reduction in the level of centralization around the July 8\textsuperscript{th} to July 13\textsuperscript{th} period are due to gaps in the source data.  Also, we have excluded the values for metrics related to account abstraction in order to get a clearer representation of the overall trends in the data.

It is important to note that the values of the master are not like other indices, insofar as it is not the case that values closer to 0 indicate more decentralization and values closer to 1 indicate more centralization.  These values are relevant in comparison the same value at other points in time.

\section{Conclusion and Discussion}

Our results clearly demonstrate that the overall Ethereum ecosystem displays elements of concentration of control that are arguably less decentralized than the community would aspire to.  The results suggest the need for diligence in efforts to develop and maintain a healthy level of decentralization.  This can seen in areas such as client diversity (in terms of network nodes), market share of staking pools, market share of rollups, and stablecoins.  While stablecoins have a much lower TVL compared to Ethereum as a whole, this belies the important function stablecoins serve in the ecosystem, and thus presents a red flag for centralization concerns.

As the ecosystem evolves, there will be changes in the important components of the ecosystem's overall infrastructure beyond the core Ethereum protocol, and there will also be changes to the extent of their impact on the overall effect level of decentralization. As Barnabé Monnot states: "There are things that the protocol does not see, but cares about" \cite{monnot2023}.  As such, we should expect our model to adapt to these changes over time.

Not only will the components that we measure change over time, but also our model will evolve as we learn more about the ecosystem and how to measure it.  While we have demonstrated that attempting to measure decentralization using a single index yields inconclusive and incomplete results, using a number of different indices together can paint a more complete picture, giving us a more holistic view of the system.

While we have attempted to use a series of metrics that can be used to cross reference each other to paint a larger picture, there are still some areas that can be explored further.  One potential area of exploration is to study the methodologies used in other areas that face a similar challenge.  One example of research that can be explored is the OECD's methodologies to measure market competition \cite{oecd2021}, in which they explore the use of numerous models and indices under two broad categories of structural measures and performance measures.

Another potential direction for research is to develop a way to measure potential collusion between entities within a population, rather than assuming that each entity in a population is totally independent, as our current model assumes.  Vitalik Buterin discussed this approach \cite{brown2023} and described it as being done by assigning pairwise coordination coefficients to all distinct pairs of entities within a population.  This would mean that any two pairwise entities within a population would have coefficient between 0 and 1, where 1 implies they are so tightly coordinated and aligned, they should be measured as a single actor, and a  0 implies they are completely independent.  The properties that attribute the level of coordination and alignment, and resultant coordination coefficient, could be as simple as being the same country, or being on the same network etc.

Obviously an approach such as this would require much more data about the individual entities in the population, and would present a challenge to gathering the data and ensuring data quality, which is an important consideration when deciding to develop a model based on that data. Nevertheless, this represents an interesting direction for future research.

One important factor in the measurement of decentralization, is that while our models become more complex, the ability for them to be widely understood remains an objective that should not be lost.  If the model itself becomes complex enough to preclude analysis from a diverse audience, then it loses its ability to act as a reference point for communication and discussion, which remains its primary objective.

\vspace{16pt}

\section{Acknowledgements}

Special thanks to the people that contributed valuable feedback and suggestions for future improvements to the model and areas for further study, including Tina Zhen, Vitalik Buterin, Stephane Gosselin and Griffin Howlett.

\newpage

\printbibliography

@inproceedings{gupta2018gini,
  title={Gini coefficient based wealth distribution in the bitcoin network: A case study},
  author={Gupta, Manas and Gupta, Parth},
  booktitle={Computing, Analytics and Networks: First International Conference, ICAN 2017, Chandigarh, India, October 27-28, 2017, Revised Selected Papers 1},
  pages={192--202},
  year={2018},
  organization={Springer}
}

@article{tran2021harnessing,
  title={Harnessing the Fifth Element of Distributional Statistics for Psychological Science: A Practical Primer and Shiny App for Measures of Statistical Inequality and Concentration},
  author={Tran, Ulrich S and Lallai, Taric and Gyimesi, Marton and Baliko, Josef and Ramazanova, Dariga and Voracek, Martin},
  journal={Frontiers in Psychology},
  volume={12},
  pages={716164},
  year={2021},
  publisher={Frontiers Media SA}
}

@inproceedings{kwon2019impossibility,
  title={Impossibility of full decentralization in permissionless blockchains},
  author={Kwon, Yujin and Liu, Jian and Kim, Minjeong and Song, Dawn and Kim, Yongdae},
  booktitle={Proceedings of the 1st ACM Conference on Advances in Financial Technologies},
  pages={110--123},
  year={2019}
}

@article{gochhayat2020measuring,
  title={Measuring decentrality in blockchain based systems},
  author={Gochhayat, Sarada Prasad and Shetty, Sachin and Mukkamala, Ravi and Foytik, Peter and Kamhoua, Georges A and Njilla, Laurent},
  journal={IEEE Access},
  volume={8},
  pages={178372--178390},
  year={2020},
  publisher={IEEE}
}

@article{karakostas2022sok,
  title={SoK: A Stratified Approach to Blockchain Decentralization},
  author={Karakostas, Dimitris and Kiayias, Aggelos and Ovezik, Christina},
  journal={arXiv preprint arXiv:2211.01291},
  year={2022}
}

@article{zhang2022sok,
  title={SoK: Blockchain Decentralization},
  author={Zhang, Luyao and Ma, Xinshi and Liu, Yulin},
  journal={arXiv preprint arXiv:2205.04256},
  year={2022}
}

@article{lee2021dq,
  title={DQ: Two approaches to measure the degree of decentralization of blockchain},
  author={Lee, Jaeseung and Lee, Byungheon and Jung, Jaeyoung and Shim, Hojun and Kim, Hwangnam},
  journal={ICT Express},
  volume={7},
  number={3},
  pages={278--282},
  year={2021},
  publisher={Elsevier}
}

@inproceedings{lin2021measuring,
  title={Measuring decentralization in bitcoin and ethereum using multiple metrics and granularities},
  author={Lin, Qinwei and Li, Chao and Zhao, Xifeng and Chen, Xianhai},
  booktitle={2021 IEEE 37th International Conference on Data Engineering Workshops (ICDEW)},
  pages={80--87},
  year={2021},
  organization={IEEE}
}

@article{sai2021characterizing,
  title={Characterizing wealth inequality in cryptocurrencies},
  author={Sai, Ashish Rajendra and Buckley, Jim and Le Gear, Andrew},
  journal={Frontiers in Blockchain},
  volume={4},
  pages={730122},
  year={2021},
  publisher={Frontiers Media SA}
}

@article{atkinson1970measurement,
  title={On the measurement of inequality},
  author={Atkinson, Anthony B and others},
  journal={Journal of economic theory},
  volume={2},
  number={3},
  pages={244--263},
  year={1970}
}

@book{rogers2010diffusion,
  title={Diffusion of innovations},
  author={Rogers, Everett M},
  year={2010},
  publisher={Simon and Schuster}
}

@article{srinivasan2018,
  author={Srinivasan, Balaji S. and Lee, Leland},
  month={6},
  title={{Quantifying Decentralization}},
  year={2018},
  url={https://news.earn.com/quantifying-decentralization-e39db233c28e},
}

@online{buterin2020,
  date={2020-10-02},
  title={A rollup-centric ethereum roadmap},
  author={Buterin, Vitalik},
  url={https://ethereum-magicians.org/t/a-rollup-centric-ethereum-roadmap/4698},
}

@inproceedings{gorzny2022ideal,
  title={Ideal properties of rollup escape hatches},
  author={Gorzny, Jan and Po-An, Lin and Derka, Martin},
  booktitle={Proceedings of the 3rd International Workshop on Distributed Infrastructure for the Common Good},
  pages={7--12},
  year={2022}
}

@misc{eigenlayer2023,
  title={{EigenLayer: The Restaking Collective}},
  author={EigenLayer Team},
  url={https://docs.eigenlayer.xyz/overview/whitepaper},
}

@misc{asgaonkar-2021,
  author={Asgaonkar, Aditya and Beekhuizen, Carl and Feist, Dankrad and Saltini, Roberto},
  month={10},
  title={{Ethereum Distributed Validator Specifications}},
  year={2021},
  url={https://github.com/ethereum/distributed-validator-specs},
}

@misc{ethereum2023,
  author={Ethereum},
  title={{Proposer-builder separation | ethereum.org}},
  url={https://ethereum.org/en/roadmap/pbs/},
}

@misc{ethereum2022,
  author={Ethereum},
  title={{Specification for the external block builders.}},
  url={https://github.com/ethereum/builder-specs},
}

@misc{labrys2022,
  author={Lebyrs},
  title={{MEV Watch}},
  url={https://www.mevwatch.info/},
}

@misc{hasu2022,
  author={Hasu},
  month={8},
  title={{Understanding liveness risks from mev-boost}},
  year={2022},
  url={https://writings.flashbots.net/understanding-mev-boost-liveness-risks},
}

@misc{gosselin2021,
  author={Gosselin, {Stephane [thegostep]}},
  month={11},
  title={{MEV-Boost: Merge ready Flashbots Architecture}},
  year={2021},
  url={https://ethresear.ch/t/mev-boost-merge-ready-flashbots-architecture/11177},
}

@misc{yang2022sok,
  title={SoK: MEV Countermeasures: Theory and Practice}, 
  author={Sen Yang and Fan Zhang and Ken Huang and Xi Chen and Youwei Yang and Feng Zhu},
  year={2022},
  eprint={2212.05111},
  archivePrefix={arXiv},
  primaryClass={cs.CR}
}

@misc{buterin2022,
  author = {Buterin, Vitalik},
  title={{State of research: increasing censorship resistance of transactions under proposer/builder separation (PBS) - HackMD}},
  url={https://notes.ethereum.org/@vbuterin/pbs_censorship_resistance},
}

@misc{buterin2021,
  author={Buterin, Vitalik},
  month={6},
  title={{Proposer/block builder separation-friendly fee market designs}},
  year={2021},
  url={https://ethresear.ch/t/proposer-block-builder-separation-friendly-fee-market-designs/9725},
}

@article{buterin2021B,
  author = {Buterin, Vitalik and Weiss, Yoav and Gazso, Kristof and Patel, Namra and Tirosh, Dror and Nacson, Shahaf and Hess, Tjaden},
  journal = {Ethereum Improvement Proposals},
  month = {9},
  number = {4337},
  title = {{ERC-4337: Account Abstraction Using Alt Mempool [DRAFT]}},
  year = {2021},
  url = {https://eips.ethereum.org/EIPS/eip-4337},
}

@article{wilson2020,
  author={Wilson, Sam and Dietrichs, Ansgar and Garnett, Matt and Zoltu, Micah},
  journal={Ethereum Improvement Proposals},
  month={10},
  title={{EIP-3074: AUTH and AUTHCALL opcodes [DRAFT]}},
  year={2020},
  url={https://eips.ethereum.org/EIPS/eip-3074},
}

@article{dietrichs2020,
  author={Dietrichs, Ansgar and Buterin, Vitalik and Garnett, Matt and Villanueva, Will and Wilson, Sam},
  journal={Ethereum Improvement Proposals},
month={9},
  title={{EIP-2938: Account Abstraction [DRAFT]}},
  year={2020},
  url={https://eips.ethereum.org/EIPS/eip-2938},
}

@article{john2023,
  author={John, Gavin},
  journal={Ethereum Improvement Proposals},
  month={1},
  title={{ERC-6315: ERC-2771 Account Abstraction [DRAFT]}},
  year={2023},
  url={https://eips.ethereum.org/EIPS/eip-6315},
}

@misc{polynya2022,
  author={Polynya},
  month={5},
  title={{Staking issuance is not net neutral}},
  year={2022},
  url={https://polynya.mirror.xyz/TpwvhuW8UsLovTAG-6EMx97cHCFBhqFs9VYgQ75qsZw},
}

@article{liu2022empirical,
  title={Empirical analysis of eip-1559: Transaction fees, waiting time, and consensus security},
  author={Liu, Yulin and Lu, Yuxuan and Nayak, Kartik and Zhang, Fan and Zhang, Luyao and Zhao, Yinhong},
  journal={arXiv preprint arXiv:2201.05574},
  year={2022}
}

@article{john2021equilibrium,
  title={Equilibrium staking levels in a proof-of-stake blockchain},
  author={John, Kose and Rivera, Thomas J and Saleh, Fahad},
  journal={Available at SSRN 3965599},
  year={2021}
}

@book{buterin2022b,
  author={Buterin, Vitalik},
  month={9},
  publisher={National Geographic Books},
  title={{Proof of Stake}},
  year={2022}
}

@article{tutberidze2018measuring,
  title={The measuring of the Gini, Theil and Atkinson indices for Georgia Republic and some other countries},
  author={TUTBERIDZE, GOCHA and PIPIA, QETEVAN and Rakviashvili, GIVI},
  journal={Globalization and Business},
  volume={5},
  pages={110--19},
  year={2018}
}

@article{gini1936measure,
  title={On the measure of concentration with special reference to income and statistics},
  author={Gini, Corrado},
  journal={Colorado College Publication, General Series},
  volume={208},
  number={1},
  pages={73--79},
  year={1936},
  publisher={Colorado College, Colorado Springs}
}

@article{kalinski2016,
  author={Kalinski, Matt},
  year={2016},
  month={04},
  pages={},
  title={The Newton-Leibnitz theorem}
}

@article{lorenz1905methods,
  title={Methods of measuring the concentration of wealth},
  author={Lorenz, Max O},
  journal={Publications of the American statistical association},
  volume={9},
  number={70},
  pages={209--219},
  year={1905},
  publisher={Taylor \& Francis}
}

@inproceedings{kusmierz2022centralized,
  title={How centralized is decentralized? Comparison of wealth distribution in coins and tokens},
  author={Kusmierz, Bartosz and Overko, Roman},
  booktitle={2022 IEEE International Conference on Omni-layer Intelligent Systems (COINS)},
  pages={1--6},
  year={2022},
  organization={IEEE}
}

@techreport{cong2023inclusion,
  title={Inclusion and democratization through web3 and defi? initial evidence from the ethereum ecosystem},
  author={Cong, Lin William and Tang, Ke and Wang, Yanxin and Zhao, Xi},
  year={2023},
  institution={National Bureau of Economic Research}
}

@article{campajola2022evolution,
  title={The evolution of centralisation on cryptocurrency platforms},
  author={Campajola, Carlo and Cristodaro, Raffaele and De Collibus, Francesco Maria and Yan, Tao and Vallarano, Nicolo' and Tessone, Claudio J},
  journal={arXiv preprint arXiv:2206.05081},
  year={2022}
}

@misc{usdoj2015,
  month={6},
  title={{Horizontal Merger Guidelines (08/19/2010)}},
  author={The United States Department of Justice},
  year={2015},
  url={https://www.justice.gov/atr/horizontal-merger-guidelines-08192010#5c},
}

@article{palma2011homogeneous,
  title={Homogeneous middles vs. heterogeneous tails, and the end of the ‘inverted-U’: It's all about the share of the rich},
  author={Palma, Jos{\'e} Gabriel},
  journal={development and Change},
  volume={42},
  number={1},
  pages={87--153},
  year={2011},
  publisher={Wiley Online Library}
}

@article{shannon1998mathematical,
  title={The Mathematical Theory of Communication; Shannon, CE, Weaver, W., Eds},
  author={Shannon, Claude E},
  journal={The mathematical theory of communication},
  pages={29--125},
  year={1998}
}

@techreport{theil1967economics,
  title={Economics and information theory},
  author={Theil, Henri},
  year={1967}
}

@article{wall2021,
  author={Wall, Eric},
  month={12},
  title={{Privacy and Cryptocurrency, Part IV: Stablecoins— Blacklists and Traceability}},
  year={2021},
  url={https://medium.com/human-rights-foundation-hrf/privacy-and-cryptocurrency-part-iv-stablecoins-for-human-rights-blacklists-and-traceability-6d74ee17c25d},
}

@book{edgington2023,
  author={Edgington, Ben},
  edition={1.3.0},
  month={4},
  title={{Upgrading Ethereum}},
  year={2023},
  url={https://eth2book.info/capella/annotated-spec/#reward-and-penalty-calculations},
}

@article{frost2023B,
  author={Frost, Jim},
  journal={Statistics By Jim},
  month={2},
  title={{Percentiles: Interpretations and Calculations}},
  year={2023},
  url={https://statisticsbyjim.com/basics/percentiles},
}

@misc{buterin2023,
  author={Buterin, Vitalik},
  month={5},
  title={{Don't overload Ethereum's consensus}},
  year={2023},
  url={https://vitalik.ca/general/2023/05/21/dont_overload.html},
}

@misc{neuder2023,
  author={Neuder, Mike and Drake, Justin},
  month={6},
  title={{Why enshrine Proposer-Builder Separation? A viable path to ePBS}},
  year={2023},
  url={https://ethresear.ch/t/why-enshrine-proposer-builder-separation-a-viable-path-to-epbs/15710/20},
}

@article{gastwirth1972estimation,
  title={The estimation of the Lorenz curve and Gini index},
  author={Gastwirth, Joseph L},
  journal={The review of economics and statistics},
  pages={306--316},
  year={1972},
  publisher={JSTOR}
}

@article{lin1991divergence,
  title={Divergence measures based on the Shannon entropy},
  author={Lin, Jianhua},
  journal={IEEE Transactions on Information theory},
  volume={37},
  number={1},
  pages={145--151},
  year={1991},
  publisher={IEEE}
}

@article{offchainlabs2023,
  author={Offchain Labs and Das, Nishant and Tsao, Terence and Van Loon, Preston and Potuz and Kirkham, Kasey and He, James},
  month={5},
  publisher={Offchain Labs},
  title={{Post-Mortem Report: Ethereum Mainnet Finality (05/11/2023)}},
  year={2023},
  url={https://offchain.medium.com/post-mortem-report-ethereum-mainnet-finality-05-11-2023-95e271dfd8b2#:~:text=Summary,and%20incurred%20an%20inactivity%20penalty.},
}

@misc{lido2023,
  title={{Lido on Ethereum Validator \& Node metrics}},
  url={https://app.hex.tech/8dedcd99-17f4-49d8-944e-4857a355b90a/app/3f7d6967-3ef6-4e69-8f7b-d02d903f045b/latest},
}

@misc{abu2023,
  author={Abu, Hanni},
  title={{Client Diversity | Ethereum}},
  url={https://clientdiversity.org/#distribution},
}

@article{checkmate2021,
  author={Checkmate},
  journal={Glassnode Insights - On-Chain Market Intelligence},
  month={5},
  title={{The Week On-chain (Week 18, 2021)}},
  year={2021},
  url={https://insights.glassnode.com/the-week-on-chain-week-18-2021/},
}

@misc{defillama,
  title={{DefiLlama}},
  url={https://defillama.com/chain/Ethereum},
}

@misc{hildobby2023,
  author={Hildobby},
  title={{ETH Stakers}},
  url={https://dune.com/queries/2394100/3928083},
  publisher={Dune Analytics}
}

@article{haqshanas2022,
  author={Haqshanas, Ruholamin},
  journal={Tokenist},
  month={8},
  title={{Tornado Cashs USDC Frozen as Stablecoin Censorship Fears Grow}},
  year={2022},
  url={https://tokenist.com/tornado-cashs-usdc-frozen-as-stablecoin-censorship-fears-grow/},
}

@article{de2022,
  author={De, Nikhilesh},
  month={2},
  title={{Circle confirms freezing 100K in USDC at law enforcements request}},
  year={2022},
  url={https://www.coindesk.com/markets/2020/07/08/circle-confirms-freezing-100k-in-usdc-at-law-enforcements-request/},
}

@misc{ethereum2023-2,
  author={Ethereum},
  title={{How The Merge impacted ETH supply | ethereum.org}},
  url={https://ethereum.org/en/roadmap/merge/issuance/},
}

@misc{ultrasoundmoney2023,
  title={{Ultra Sound money}},
  url={https://ultrasound.money/},
}

@misc{beaconchain2023,
  title={{Open Source Ethereum Blockchain Explorer - Beaconcha.in - 2023}},
  url={https://beaconcha.in/},
}

@misc{etherscan2023,
  title={{Ether Supply Growth Chart}},
  url={https://etherscan.io/chart/ethersupplygrowth},
}

@article{monnot2023,
  author={Monnot, Barnabé},
  month={4},
  title={{Seeing like a protocol}},
  year={2023},
  url={https://barnabe.substack.com/p/seeing-like-a-protocol},
}

@article{oecd2021,
  author={OECD},
  journal={OECD Competition Committee Issues},
  month={6},
  title={{Methodologies to measure market competition}},
  year={2021},
  url={https://www.oecd.org/daf/competition/methodologies-to-measure-market-competition.htm},
}

@misc{brown2023,
  author={Simon Brown},
  month={7},
  title={{Simon Brown - Presentation @ PBS Salon :: ETHCC 2023}},
  year={2023},
  url={https://www.youtube.com/watch?v=FUzcN3Mp-6E},
}

\newpage

\onecolumn

\section{Appendix: List of Data Sources}

\vspace{12pt}

The following table lists the various sources of data for each data point that is referenced in our model.  The data from these sources were recorded programmatically and compiled into a database using software that was specifically designed for that purpose.

\vspace{24pt}

\begin{table*}[ht]
\normalsize
\begin{tabular}{ll}
\multicolumn{2}{l}{\textbf{Based on original Nakamoto Coefficient subsystem selection:}} \\[10pt]
Consensus nodes by client & \url{https://migalabs.es/api/v1/client-distribution} \\[6pt]
Consensus nodes by country & \url{https://migalabs.es/api/v1/geo-distribution} \\[6pt]
Execution nodes by client & \url{https://www.ethernodes.org/} \\[6pt]
Execution nodes by country & \url{https://www.ethernodes.org/countries} \\[6pt]
Distribution of native asset by amount & \url{https://data.messari.io/api/v1/assets/ethereum/metrics} \\[6pt]
Amount staked by Pool / Staking Service Provider & \url{https://api.dune.com/api/v1/query/2394100/results} \\[24pt]
\multicolumn{2}{l}{\textbf{Metrics pertaining to Proposer Builder Separation:}} \\[10pt]
Blocks proposed by builder & \url{https://mevboost.pics/data.html} \\[6pt]
Blocks proposed by relayer & \url{https://mevboost.pics/data.html} \\[24pt]
\multicolumn{2}{l}{\textbf{Metrics pertaining to Account Abstraction:}} \\[10pt]
Number of user operations per bundler & \url{https://dune.com/queries/2193933/3599135} \\[6pt]
Number of wallets per deployer & \url{https://dune.com/queries/2434102/3999582} \\[24pt]
\multicolumn{2}{l}{\textbf{Miscellaneous Metrics:}} \\[10pt]
Rollups by relative TVL / size / volume & \url{https://l2beat.com/scaling/tvl} \\[6pt]
Stablecoins & \url{https://stablecoins.llama.fi/stablecoins} \\[6pt]
\end{tabular}
\end{table*}

\end{document}